\documentclass[a4paper,11pt]{article}
\pdfoutput=1 % if your are submitting a pdflatex (i.e. if you have
\usepackage{jheppub} % for details on the use of the package, please
\usepackage[T1]{fontenc} % if needed

\title{High precision measurement of cosmic curvature: from gravitational waves and cosmic chronometer}

\author[a]{Y. He,}
\author[a,c,1]{Y. Pan,\note{Corresponding author.}}
\author[b,1]{D-P Shi,\note{Corresponding author.}}
\author[c]{J. Li}
\author[d]{S. Cao}
\author[a,c]{W. Cheng}
\affiliation[a]{Chongqing University of Posts and Telecommunications,\\400065, Chongqing, China}
\affiliation[b]{School of Electronic and Electrical Engineering, Chongqing University of Arts and Science, \\402160, Chongqing, China}
\affiliation[c]{Chongqing University,\\400044, Chongqing, China}
\affiliation[d]{Departmentof Astronomy, Beijing Normal University, Beijing,\\100875,China}

\emailAdd{s190601005@stu.cqupt.edu.cn}
\emailAdd{panyu@cqupt.edu.cn}
\emailAdd{dpshi@cqwu.edn.cn}
\emailAdd{cqujinli1983@cqu.edu.cn}
\emailAdd{caoshuo@bnu.edu.cn}
\emailAdd{chengwei@cqupt.edu.cn}

\abstract{Although the spatial curvature has been measured with very high precision, it still suffers from the well known cosmic curvature tension. In this paper, we propose an improved method to determine the cosmic curvature, by using the simulated data of binary neutron star mergers observed by the second generation space-based DECi-hertz Interferometer Gravitational-wave Observatory (DECIGO). By
applying the Hubble parameter observations of cosmic chronometers to the DECIGO standard sirens, we explore different possibilities of making measurements of the cosmic curvature referring to a distant past: one is to reconstruct the Hubble parameters through the Gaussian process without the influence of hypothetical models, and the other is deriving constraints on $\Omega_K$ in the framework of non-flat $\Lambda$ cold dark matter model. It is shown that in the improved method DECIGO could provide a reliable and stringent constraint on the cosmic curvature ($\Omega_{K} = -0.007\pm0.016$), while we could only expect the zero cosmic curvature to be established at the precision of $\Delta \Omega_K=0.12$ in the second model-dependent method. Therefore, our results indicate that
in the framework of methodology proposed in this paper, the increasing number of well-measured standard sirens in DECIGO could significantly reduce the bias of estimations for cosmic curvature. Such constraint is also comparable to the precision of Planck 2018 results with the newest cosmic microwave background (CMB) observations ($\Delta \Omega_{K} \approx 0.018$), based on the concordance $\Lambda$CDM model.}


\begin{document}
\maketitle
\flushbottom

\section{Introduction}
\label{sec:intro}

The determination of cosmic curvature is an important research in cosmology. As we all known, satisfying the principles of cosmology, the space-time of our universe can be described by the Friedmann-Lema\^itre-Robertson-Walker metric. Cosmic curvature can help us understand whether the space of universe is open, closed or flat. It is worth mentioning that in the research of some scholars, the cosmic curvature is not completely independent, and there is a certain dependence between dark energy and cosmic curvature\citep{ Clarkson2007,Wang2007,Hlozek2008}.

In order to determine the cosmic curvature, many scholars have carried out related research. Scholars mainly study curvature in two aspects. On the one hand, based on the model assumption of the cold dark matter model, it is feasible to fix the dark energy parameters in the model and introduce the cosmic curvature density parameter. In these research, the constraint result of the curvature density parameter by the Planck2018 cosmic microwave background (CMB) is $-0.095<\Omega_{K}<-0.007$\citep{Aghanim2020,DiValentino2020}. The constraints of CMB combined with lens and baryon acoustic oscillation(BAO) on the parameter of cosmic curvature density support the flat universe, $\Omega_{K}=0.007~\pm 0.0019$\citep{Aghanim2020}. Besides, Gao et al.\cite{Gao2020} studied the curvature density parameters and dark energy using the latest supernova samples. The study of using full-shape and BAO to constrain the model with curvature parameters has also been proposed\citep{Chudaykin2021}. On the other hand, progress has been made in measuring curvature using model-independent methods. Starting from the equation of curvature parameters, we can construct the curvature parameters at different redshifts by providing Hubble parameters($H(z)$) and luminosity distances($D_{L}$) at different redshifts $\Omega_{K} = \frac{(H(z)D^{\prime}(z))^2-c^2} {H_{0}^{2}D(z)^2}$ \citep{Clarkson2007}. This method provides a good way to directly test the curvature parameters by bypassing the hypothesis of the model. Li et al. used $H(z)$ and angular diameter distance($D_A(z)$) given by BAO to test the curvature parameters\cite{Li2014}. Supernova data and quasar data, which do not depend on model assumptions, are also suitable for these works\citep{Gao2020,Yahya2014,Cai2016,Li2016,Wang2017,Liu2020b,Jesus2021,Cao2019b,Wei2020}. Some scholars use gravitational waves and $H(z)$ to construct curvature at different redshifts\cite{Zheng2021}.
It is worth noting that this method needs to estimate the first and second derivatives of the luminosity distance from the fitting function, which will lead to the increase of uncertainty\cite{Yu2016}. Therefore, for the deficiency of this method, some scholars have proposed proposed an improved model independent method to measure curvature\citep{Wei2017}. In their research, the error caused by the derivative of luminosity distance is overcome. Since then, some scholars have used this method to test the curvature of the universe\cite{Wei2018,Cao2019b,Wei2020,Yang2021}. Many scholars use model-independent methods to test the curvature parameters. For example, based on the distance ratio contained in the strong gravitational lensing, some scholars avoid assuming the model to test the curvature\citep{Liu2020a,Wang2020,Zhou2020,Qi2021}. In addition, some scholars have tested the curvature using strong lens time delays data\cite{Liao2017}. It should be noted that although cosmological models do not need to be assumed in these methods, the Gaussian process used to reconstruct data may have a model dependence tendency\cite{Colgain2021GP}. It is worth mentioning that recently, some scholars have used machine learning methods combined with $H(z)$ and supernova samples to test the curvature parameters $\Omega_{K} = 0.028\pm 0.186$ \cite{Wang2021}. In their research, the method based on artificial neural network can overcome the a priori problem of Hubble constant($H_{0}$).

The successful detection of gravitational waves(GW) opens a new window for cosmological research\citep{Abbott2016,Abbott2017}. Like supernovae, GW do not rely on model assumptions in detection. Looking for merging events of binary black holes or merging events of two neutron stars with electromagnetic radiation can be used as standard sirens to study cosmology\citep{Dalal2006,Hlozek2008,Zhao2011,Liao2017NC,Cai2017,Du2022}. Recently, black hole-neutron star merging events have also been successfully detected \citep{Abbott2021}. Moreover, the coupling coefficient between GW and matter is small, which can carry more primordial wave source information, which benefits of the cosmology. In this paper, the data from the space gravitational wave detector DECi-hertz Interferometer Gravitational wave Observatory (DECIGO) are used to simulate the gravitational wave events that will be detected by DECIGO in the future. And we use two methods to test curvature. In the first method, we directly test the curvature density parameters in the luminosity distance, but the cosmological model is not added to the luminosity distance. In the second method, we add a cosmological model including curvature to the luminosity distance, and constrain the curvature density parameters in the model.
First, we use the improved method proposed Wei et al.\citep{Wei2017} to test the curvature. And taking into account the impact of the number of data samples on the constraints, we reconstruct the expansion rate, and the curvature parameters are constrained. Then, we use gravitational wave data to constrain $\Omega_K$ in the framework of non-flat $\Lambda$ cold dark matter model($\Lambda CDM$+$\Omega_K$ model) to compare the curvature constraint effect of the two methods. In addition, in the second method, we use electromagnetic wave(EM) (supernovae and quasars) data to constrain the curvature in $\Lambda CDM+\Omega_{K}$ model to compare with the constraints from DECIGO GW data. Finally, we compare and analyze the results with the research of other scholars.

The structure of this paper is as follows. In Sec. \ref{sec:DAM}, we briefly introduce the data and methods used. In Sec. \ref{Rsu}, we will give the results and analysis. In Sec. \ref{Con}, we will make a summary of this article.

\section{Data and Method} \label{sec:DAM}

\subsection{Gravitational Wave Detection From DECIGO}
For GW events, we know that the luminosity distance of the merging event of two stars can be obtained by detection, and the redshift information can be obtained by directly detecting the merging event with neutron stars\citep{Holz2005,Zhao2011,Abbott2017}. So this kind of binary merging events can be used as a standard siren to study cosmology.

DECIGO\citep{Seto2001,Kawamura2006} is a GW detection project under construction in Japan, which means DECi-hertz Interferometer Gravitational wave Observatory. DECIGO's detection frequency ranges from 0.1Hz to 10Hz\citep{Kawamura2019}. In this paper, we use DECIGO as the detector to simulate the standard siren information provided by gravitational waves.

Consider two systems with masses $m_1$ and $m_2$, whose Fourier transform can be expressed as \\

\begin{equation}
\tilde{h}(f)=\frac{A}{d_{L}(z)} M_{z}^{5 / 6} f^{-7 / 6} e^{i \Psi(f)} ,
\end{equation}

Where, $\Psi(f)$ and $A=\left(\sqrt{6} \pi^{2 / 3}\right)^{-1}$ are the inspiral phase term and geometrical average over the inclination angle of the system respectively. The former parameter, $\Psi(f)$, is removed from the calculation, which is a function of the coalescence time $t_c$, while the latter is a constant. $d_{L}(z)$ is a function of luminosity distance with respect to $z$ and $M_{z}=(1+z) \eta^{3 / 5} (m_1 +m_2)$ is the redshifted chirp mass. $\eta=m_{1} m_{2} / (m_1 + m_2)^{2}$ is the symmetric mass ratio.According to the studies of other scholars\citep{Sathyaprakash2010,Zhao2011}, we assume that the neutron stars are uniformly distributed at [1,2]$M_{\odot}$, and the coalescence time and initial phase of emission are both zero(i.e. $\left(t_{c}=0, \phi_{c}\right)=0$). In this way, the unknown parameters can be reduced to three: $\theta=\left\{M_{z}, \eta, d_{L}\right\}$.

We use the fisher matrix to estimate the uncertainty:

\begin{equation}
\Gamma_{a b}=4 R e \int_{f_{\min }}^{f_{\max }} \frac{\partial_{a} \tilde{h}_{i}^{*}(f) \partial_{b} \tilde{h}_{i}(f)}{S_{h}(f)} d f ,
\end{equation}

Where $\partial_{a}$ means to derive the parameter $\theta_a$. It can be seen from the data released by DECIGO that it has eight equivalent detectors\citep{Seto2001,Kawamura2006}. Therefore, if all detectors are taken into account, the coefficient of $\Gamma_{a b}$ should be 8 times that of a single detector.Noise spectrum Analysis from DECIGO\citep{Kawamura2006, Nishizawa2010, Kawamura2019}:

\begin{eqnarray}
S_{h}(f) & = &6.53 \times 10^{-49}\left[1+\left(\frac{f}{7.36 H z}\right)^{2}\right] \nonumber \\
&&+ 4.45 \times 10^{-51} \times\left(\frac{f}{1 H z}\right)^{-4} \times \frac{1}{1+\left(\frac{f}{7.36 H z}\right)^{2}} \nonumber \\
&&+ 4.94 \times 10^{-52} \times\left(\frac{f}{1 H z}\right)^{-4} \mathrm{~Hz}^{-1} ,
\end{eqnarray}

Where the first line on the right represents shot noise, the second line represents radiation pressure noise, and the last line represents acceleration noise.

In the process of simulating GW, we need to assume the parameters of the cosmological model. In this paper, based on the data given by Bennett et al.\citep{Bennett2014}, we choose the $\Lambda CDM$ model of ($H_0=69.6 km~s^{-1}~Mpc^{-1}, \Omega_{m} = 0.286$). Luminosity distance given by simulated GW

\begin{equation}
d_{L, G W}(z)=\frac{c(1+z)}{H_{0}} \int_{0}^{z} \frac{d z^{\prime}}{\sqrt{\Omega_{m}\left(1+z^{\prime}\right)^{3}+\left(1-\Omega_{m}\right)}} ,
\end{equation}

Where c is the speed of light and $H_0$ is the Hubble constant.

The distribution of wave sources that can be observed on Earth is\citep{Sathyaprakash2010,Cai2017}:

\begin{equation}
P(z) \propto \frac{4 \pi D_{C}^{2}(z) R(z)}{H(z)(1+z)} ,
\end{equation}

Where, $H(z)$ represents the Hubble parameter, $D_C(z)$ represents the comoving distance, and the representation of $R(z)$ has been used in many articles\citep{Schneider2001, Cutler2006,Cai2017}

\begin{equation}
R(z)=\left\{\begin{array}{ll}
1+2 z, & z \leq 1 \\
\frac{3}{4}(5-z), & 1<z<5 \\
0, & z \geq 5
\end{array}\right. ,
\end{equation}

The error expression of luminosity distance is:

\begin{eqnarray}
\sigma_{d_{L, G W}} &=&\sqrt{\left(\sigma_{d_{L, G W}^{\mathrm{inst}}}\right)^{2}+\left(\sigma_{D_{d, G W}^{\mathrm{lens}}}\right)^{2}} \\
&=&\sqrt{\left(\frac{2 d_{L, G W}}{\rho}\right)^{2}+\left(0.05 z d_{L, G W}\right)^{2}} ,
\end{eqnarray}

Where, $d_{L, G W}^{\mathrm{inst}}$ represents the error caused by noise, $d_{L, G W}^{\mathrm{lens}}$ represents the error caused by weak gravitational lens effect. In the above equation, the extension of the second equation comes from \cite{Zhao2011} and \cite{Sathyaprakash2010}, respectively. $\rho$ represents signal-to-noise ratio.

\begin{figure}[ht!]
\centering % \begin{center}/\end{center} takes some additional vertical space
\includegraphics[width=.45\textwidth,clip]{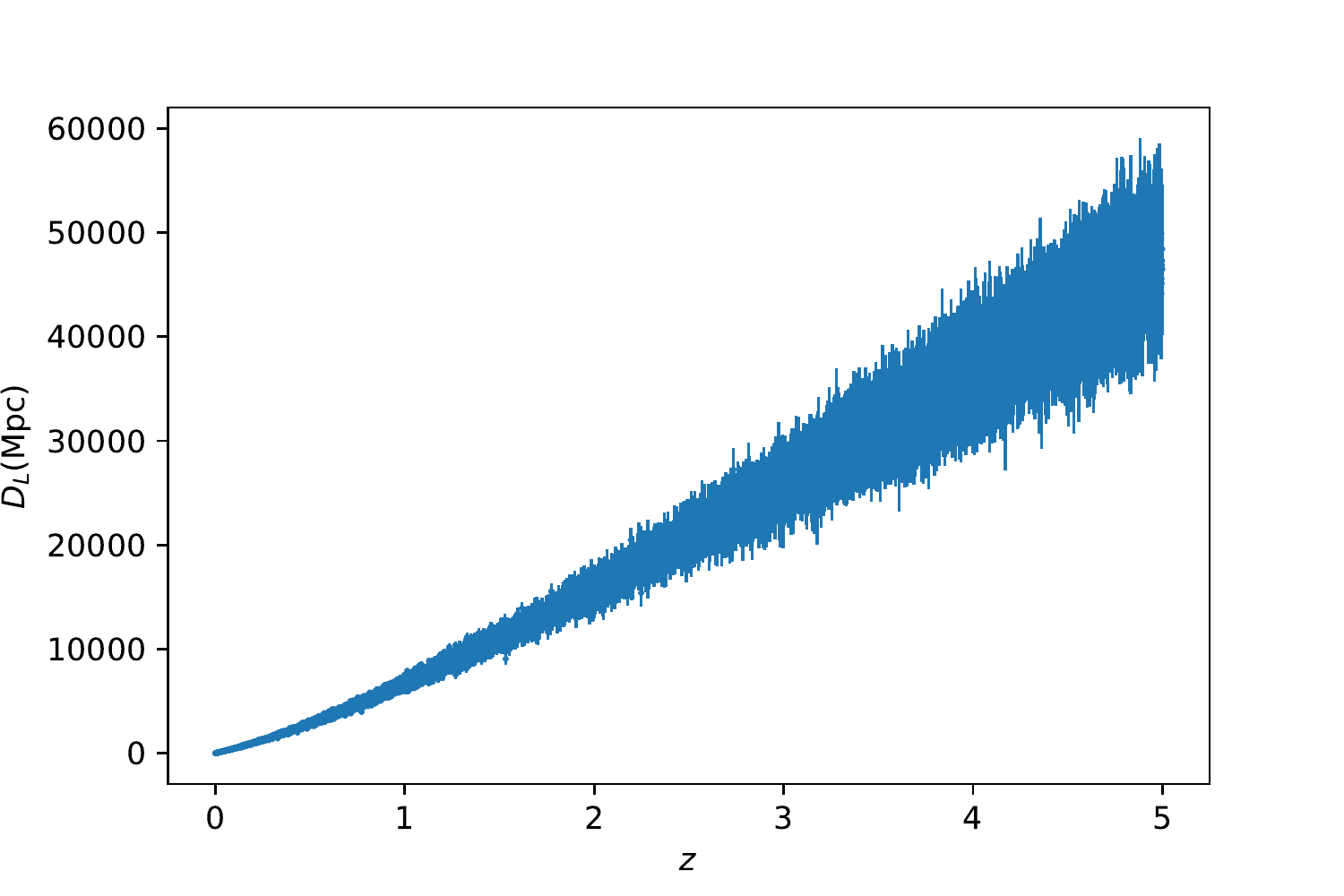}
\caption{\label{fig:DECIGODL} Luminosity distance and redshift data simulated by 10000 GW events.}
\hfill
\end{figure}

Through the analysis of \cite{Kawamura2019}, DECIGO expects to detect more than 10000 binary merging events every year, and based on the analysis of \cite{Cutler2009}, it is feasible to determine the redshift of these events through their electromagnetic counterparts. Therefore, we simulate the luminosity distance and the corresponding redshift of 10000 GW events, see fig. \ref{fig:DECIGODL}.

\subsection{Gaussian Process} \label{sec:GPHz}
In this paper, in order to calculate the data at different redshifts at the same redshift, we use the Gaussian process method to reconstruct the data. This method was first used by Seikel et al.\citep{Seikel2012}, and has been studied by many scholars\citep{Zhang2018,Liao2019,Fu2013Yu,Li2016Yu,Liu2020b,Zhou2020}.

Gaussian process is a method of smoothing data. For the input data set, multiple smooth data in a given range can be reconstructed, and the tasks of data set expansion and redshift reconstruction can be well completed. Seikel et al.\citep{Seikel2012} used GP to reconstruct the dark energy state parameter equation, which includes the use of GP, for which they have developed a third-party library GaPP based on Python.

In this paper, we use Gaussian process to reconstruct Hubble parameter data. The evolution of Hubble parameter with redshift represents the change of cosmic expansion rate with the increase of distance. Jimenez et al.\citep{Jimenez2002} pointed out that we can use the relative galactic age to constrain the cosmological parameters, in their study, the Hubble parameter is expressed as

\begin{equation}
H(z)=-\frac{1}{(1+z)} \frac{d z}{d t} ,
\end{equation}

In this work, we use 31 redshifts from 0.09 to 1.965 $H(z)$ data from cosmic-chronometer approach\citep{Jimenez2003,Simon2005,Stern2010,Moresco2012,Moresco2015,Moresco2016,Ratsimbazafy2017}.

\begin{figure}[ht!]
\centering % \begin{center}/\end{center} takes some additional vertical space
\includegraphics[width=.45\textwidth,clip]{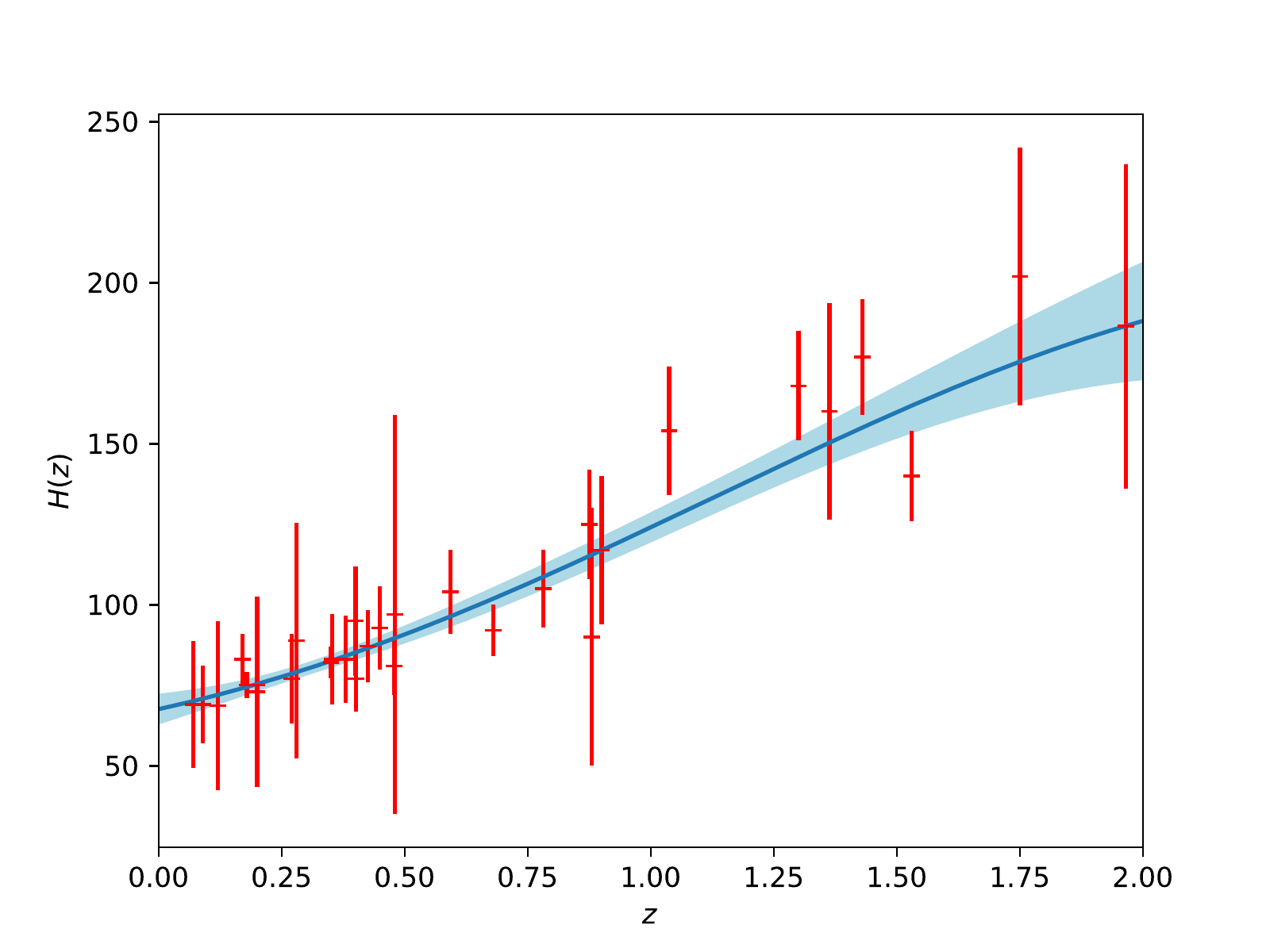}
\caption{\label{fig:Hz} We reconstruct 31 Hubble parameter data from the cosmic chronometer and show its error within 1 $\sigma$.}
\hfill
\end{figure}

\begin{figure}[ht!]
\centering % \begin{center}/\end{center} takes some additional vertical space
\includegraphics[width=.45\textwidth,clip]{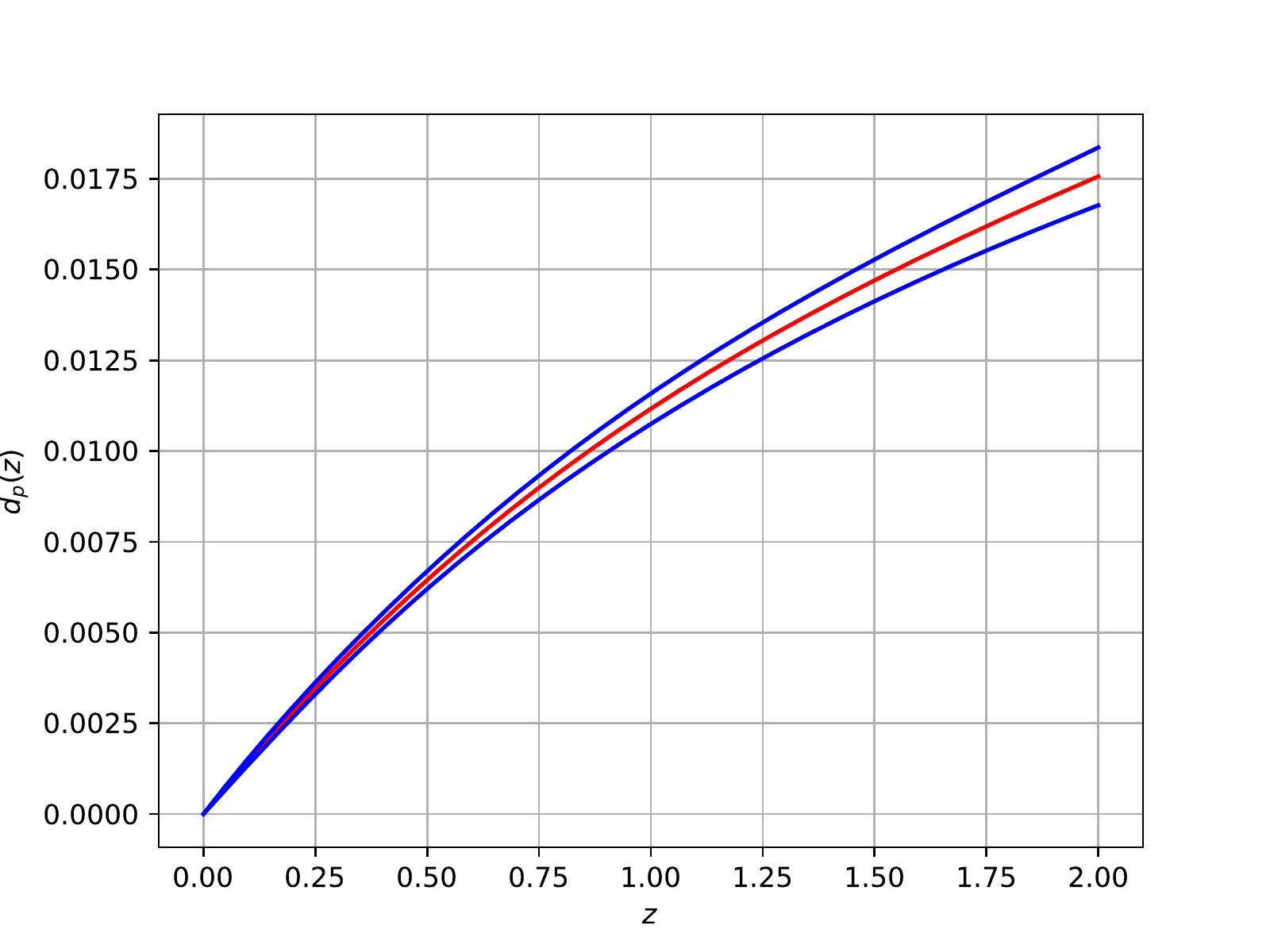}
\caption{\label{fig:dpz} We get $d_p(z)$ and its error for the reconstructed Hz integral, and the red line represents the error of $d_p(z)$ within 1 $\sigma$.}
\hfill
\end{figure}

We reconstruct 31 Hubble parameter data from the cosmic chronometer, and the results are shown in fig. \ref{fig:Hz}. In addition, we also integrate the reconstruction results and get the $d_{p}(z)$ and its error within 1 $\sigma$, as shown in fig. \ref{fig:dpz}.

\subsection{Luminosity Distance From The Cosmic Chronometer}

As we all know, in the Friedmann-Lemaitre-Robertson-Walker(FLRW) metric, the luminosity distance can be expressed as:

\begin{eqnarray}\label{eq1}
d_{L}(z)=\left\{\begin{array}{ll}
\frac{c(1+z)}{H_{0} \sqrt{\left|\Omega_{\mathrm{K}}\right|}} \sinh \left[\sqrt{\left|\Omega_{\mathrm{K}}\right|} \int_{0}^{z} \frac{d z^{\prime}}{E\left(z^{\prime}\right)}\right], &  \Omega_{K}>0 \\
\frac{c(1+z)}{H_{0}} \int_{0}^{z} \frac{d z^{\prime}}{E\left(z^{\prime}\right)}, &  \Omega_{K}=0. \\
\frac{c(1+z)}{H_{0} \sqrt{\left|\Omega_{\mathrm{K}}\right|}} \sin \left[\sqrt{\left|\Omega_{\mathrm{K}}\right|} \int_{0}^{z} \frac{d z^{\prime}}{E\left(z^{\prime}\right)}\right], &  \Omega_{K}<0
\end{array}\right. \nonumber \\
\end{eqnarray}

$\Omega_{K}$ represents the curvature density parameter and $z$ represents the redshift. $E(z)=H(z)/H_0$ is given by the ratio of $H(z)$ to $H_0$. $H(z)$(Hubble parameter) denotes the expansion rate at $z$, and $H_0$ is the Hubble parameter at $z=0$, which is called the Hubble constant. In order to eliminate the influence of the model, we can introduce comoving distance:

\begin{equation}\label{eq2}
d_{p}(z)=\int_{0}^{z} \frac{d z^{\prime}}{H\left(z^{\prime}\right)} .
\end{equation}

If we express $d_p(z)$ in terms of the Hubble parameter, we can get a luminosity distance given by the Hubble parameter,

\begin{eqnarray}\label{eq3}
d_{L}^{Hz}(z)=\left\{\begin{array}{ll}
\frac{c(1+z)}{H_{0} \sqrt{\left|\Omega_{\mathrm{K}}\right|}} \sinh \left[\sqrt{\left|\Omega_{\mathrm{K}}\right|} H_{0} d_p(z)\right], &  \Omega_{K}>0 \\
\frac{c(1+z)}{H_{0}} H_{0} d_p(z), &  \Omega_{K}=0. \\
\frac{c(1+z)}{H_{0} \sqrt{\left|\Omega_{\mathrm{K}}\right|}} \sin \left[\sqrt{\left|\Omega_{\mathrm{K}}\right|} H_{0} d_p(z)\right], &  \Omega_{K}<0
\end{array}\right. \nonumber \\
\end{eqnarray}

Accordingly, the error of $d_{L}^{Hz}(z)$ is given by error transmission,

\begin{eqnarray}\label{eq4}
\sigma_{d_{L}^{Hz}(z)}=\left\{\begin{array}{ll}
\frac{c(1+z)}{H_{0}} \cosh \left[\sqrt{\left|\Omega_{K}\right|}  H_{0} d_p(z) \right] \sigma_{d_p(z)}, & \Omega_{K}>0 \\
c(1+z) \sigma_{d_p(z)}, & \Omega_{K}=0 \\
\frac{c(1+z)}{H_{0}} \cos \left[\sqrt{\left|\Omega_{K}\right|} H_{0} d_p(z) \right] \sigma_{d_p(z)}, & \Omega_{K}<0
\end{array}\right. \nonumber \\
\end{eqnarray}

In this paper, we set the Hubble constant to $H_0=69.6\pm0.7~km~s^{-1}Mpc^{-1}$ \cite{Bennett2014}, so the only free parameter contained in the $d_{L}^{Hz}(z)$ is $\Omega_{K}$. We will use GW data to constrain the parameter $\Omega_K$ through Markov chain Monte Carlo method in Sec. \ref{Rsu}.

\section{Result and Discussion} \label{Rsu}
\subsection{Improved Curvature Test Method}
In the \ref{sec:GPHz}, we use the Gaussian process to reconstruct the Hubble parameter data and obtain the luminosity distance($d_{L}^{Hz}(z)$) given by $H(z)$. It is worth noting that the luminosity distance obtained by this method is not exactly the same as that given by GW, so we smooth the $d_{L}^{Hz}(z)$. Considering that there may be unknown errors in gravitational wave detection, we add 10\% systematic error to $d_{L}^{GW}(z)$. Therefore, we can constrain the curvature density parameter through $\chi^2$ method:

\begin{equation}
\chi^{2} = \sum_{i} \frac{\left[d_{L}^{Hz}\left(z_{i} ; \Omega_{K}\right)-d_{L}^{\mathrm{GW}}\left(z_{i}\right)\right]^{2}}{\sigma_{d_{L}^{Hz}, i}^{2}+\sigma_{d_{L}^{\mathrm{GW}}, i}^{2}} ,
\end{equation}

\begin{figure}[ht!]
\centering % \begin{center}/\end{center} takes some additional vertical space
\includegraphics[width=.45\textwidth,clip]{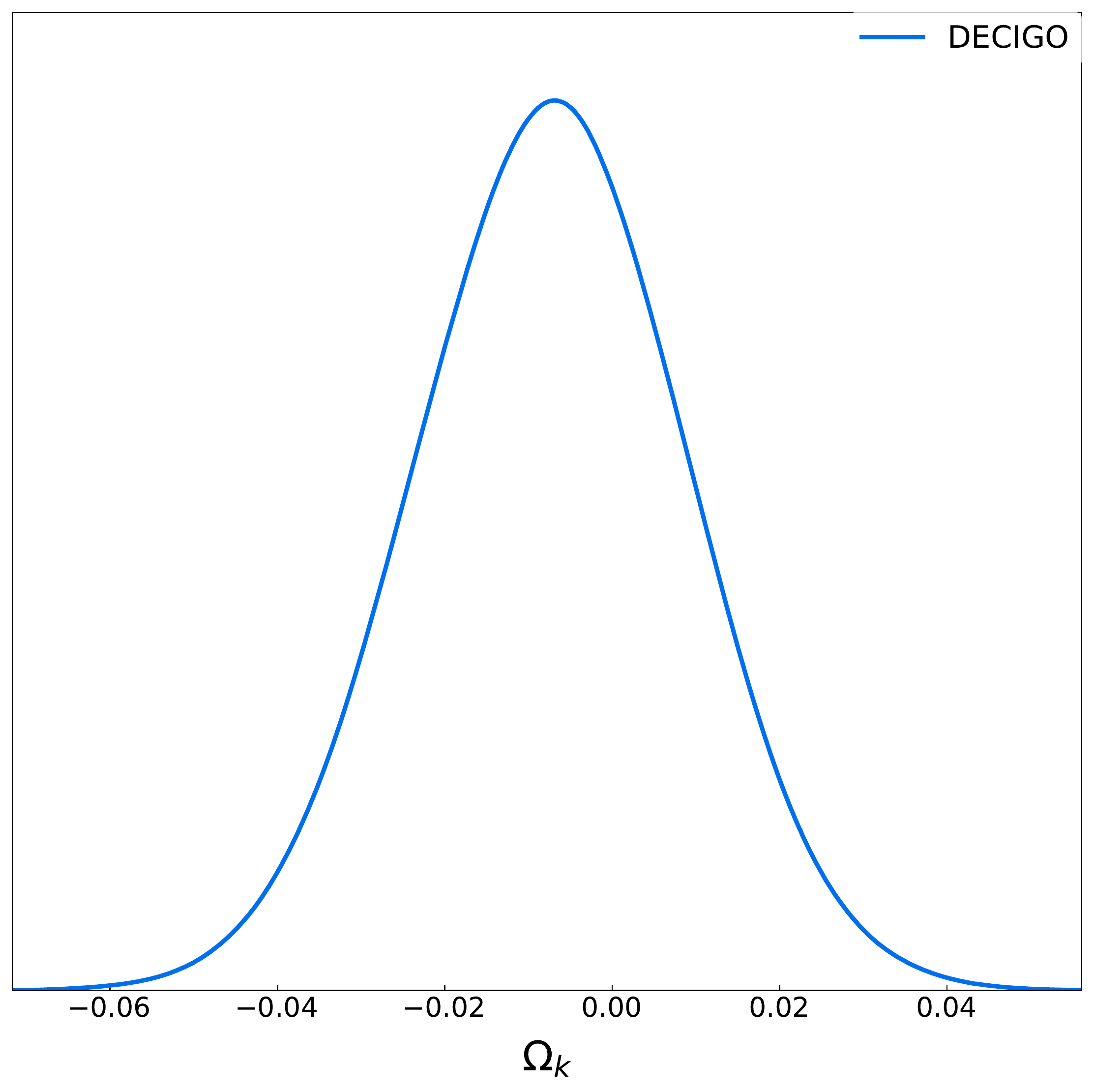}
\caption{\label{fig:DECIGOOK} The constraint results of the reconstructed cosmic chronometer and DECIGO on the curvature parameter.}
\hfill
\end{figure}

The curvature parameter result is shown in Fig.\ref{fig:DECIGOOK}, and in order to show the constraint effect more intuitively, we compare the constraint results of curvature parameters given by other observation data using the same method (see Tab.\ref{tab:Omegak1}).

\begin{figure}[ht!]
\centering % \begin{center}/\end{center} takes some additional vertical space
\includegraphics[width=.45\textwidth,clip]{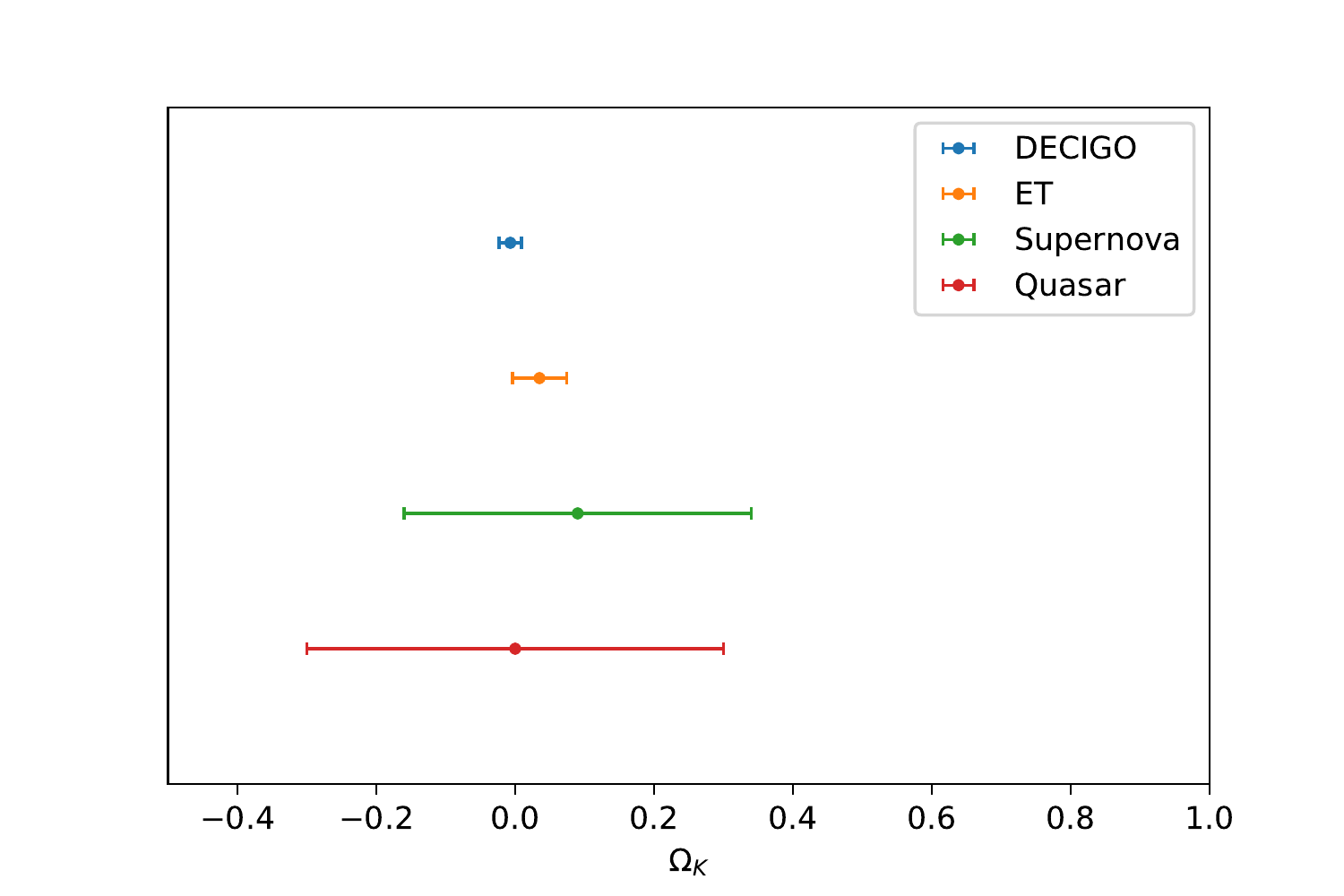}
\caption{\label{fig:fourData} Comparison of the results of four kinds of data using model-independent method.}
\hfill
\end{figure}

As can be seen from the Tab. \ref{tab:Omegak1}, the cosmic curvature density parameter which is constrained by the luminosity distance obtained from the reconstructed $H(z)$ and the simulated GW data from DECIGO is $\Omega_{K} = -0.007\pm0.016$. In order to better compare the constraint ability of DECIGO to curvature parameters, we add the research results of other scholars to the table. The constraint result given by the third generation gravitational wave detector ET is $\Omega_{K} = 0.035\pm0.039$, the curvature constraint ability of DECIGO is better than that of ET.
We also show the curvature constraint results of supernovae and quasars. It is easy to see that the curvature constraint accuracy given by the GW data is more than 90\% higher than that given by quasars ($\Omega_{K} = 0.0\pm0.3$) and supernovae data ($\Omega_{K} = 0.09\pm0.25$) respectively.
In addition, we also noticed that Zheng et al.\citep{Zheng2021} used the GW simulation data of DECIGO and ET to test the cosmic curvature, and they used the third-order logarithm polynomial approximation of $D_{L}(z)$ with undetermined coefficients in their research, and directly constrained the cosmic curvature through the research results of Clarkson et al.\citep{Clarkson2007}. They use DECIGO(0<z<2) to simulate data to constrain curvature to $\Omega_{K} = 0.004\pm0.09$, and the result of using ET data in the same way is $\Omega_{K} = 0.01\pm0.10$. Compared with the results of Zheng et al.\citep{Zheng2021}, the research method used in this paper gives higher constraint accuracy. As the space gravitational wave detector, the accuracy of DECIGO is more than 50\% higher than that of ET detector when using the same method to measure curvature density parameters. In order to more intuitively compare the results given by different data, we visualize the comparison of the data (see Fig. \ref{fig:fourData}).

\begin{table}[tbp]
\centering
\begin{tabular}{c|cccccc}
\hline
Data            & $\Omega_{K}$      &         &Source\\
\hline
DECIGO          & $-0.007\pm0.016$&         &This~Work\\
ET              & $0.035\pm0.039$ &         &\citep{Wei2018}\\
Quasar          & $0.0\pm0.3$     &         &\citep{Cao2019b}\\
supernove       & $0.09\pm0.25$   &         &\citep{Wei2017}\\
\hline
\end{tabular}
\caption{\label{tab:Omegak1} The results obtained by using the model-independent method to constrain curvature.}
\end{table}

\subsection{Curvature Test from Cosmological Model}

In addition, we also investigate the constraint effect of DECIGO on curvature density parameters under the assumption of cosmological model.
The $\Lambda CDM+\Omega_{K}$ model is,

\begin{equation}
E(z) =  \Omega_m (1+z)^{3} +\Omega_K(1+z)^2 + (1-\Omega_m-\Omega_K),
\end{equation}

At the same time, we also combine the supernova\cite{Scolnic2018} data with quasar\cite{Cao2017a,Cao2017b} data (EM) to constrain curvature parameter to discuss the constraint effect of GW data and EM data on the curvature density parameters in the $\Lambda CDM+\Omega_{K}$ model. Meanwhile, we divide the sub-samples of simulated GW with redshifts $0<z<2$ from the full samples with redshifts $0<z<5$ and assess the accuracy of $\Omega_{K}$ given by the model-dependent($\Lambda CDM+\Omega_{K}$ model) method and Improved curvature test method.

\begin{figure}[ht!]
\centering % \begin{center}/\end{center} takes some additional vertical space
\includegraphics[width=.45\textwidth,clip]{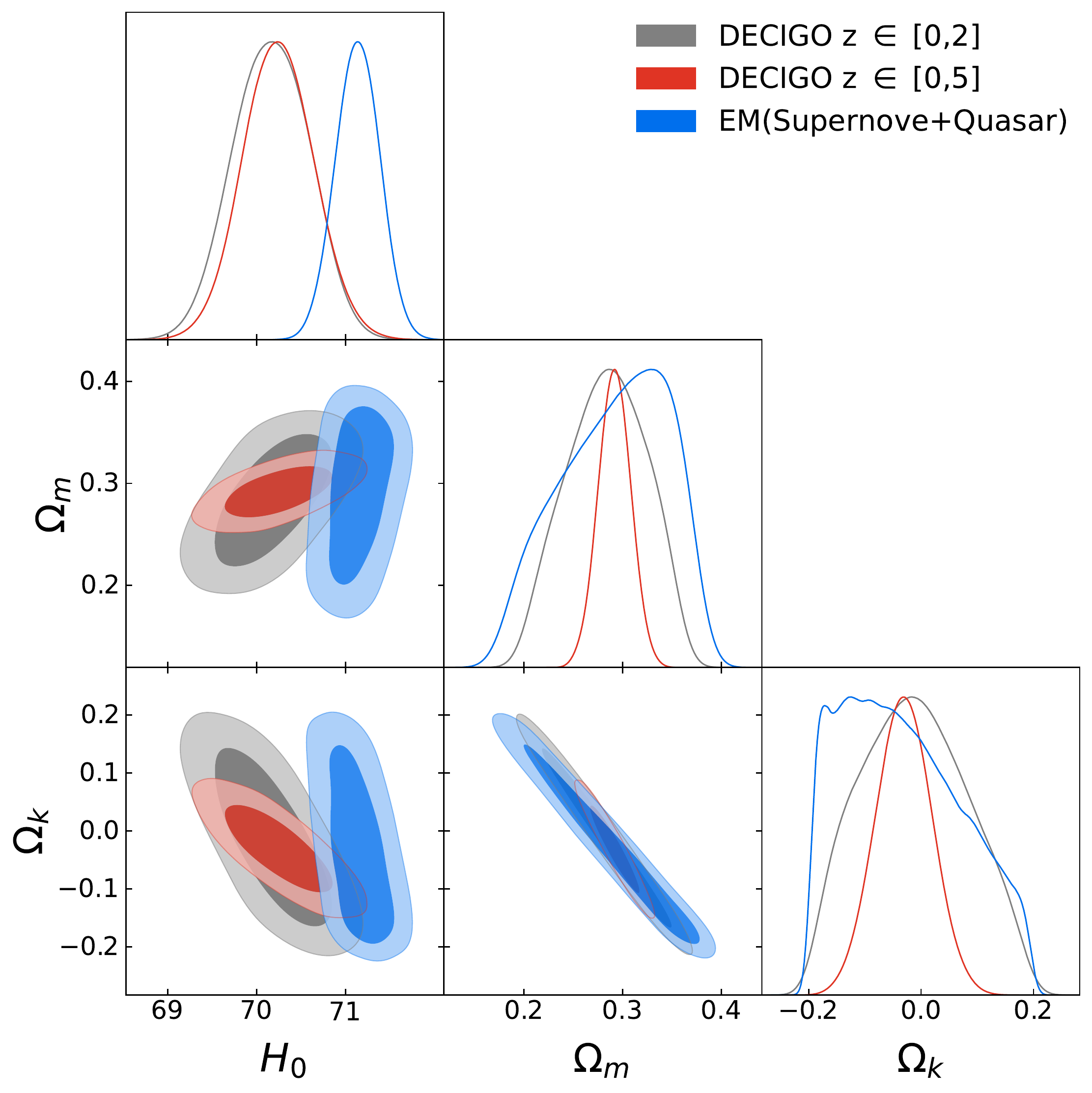}
\caption{\label{fig:DESNQUOk} The result of constraining the $\Lambda CDM+\Omega_{K}$ model using DECIGO and EM, respectively.}
\hfill
\end{figure}

\begin{figure}[ht!]
\centering % \begin{center}/\end{center} takes some additional vertical space
\includegraphics[width=.45\textwidth,clip]{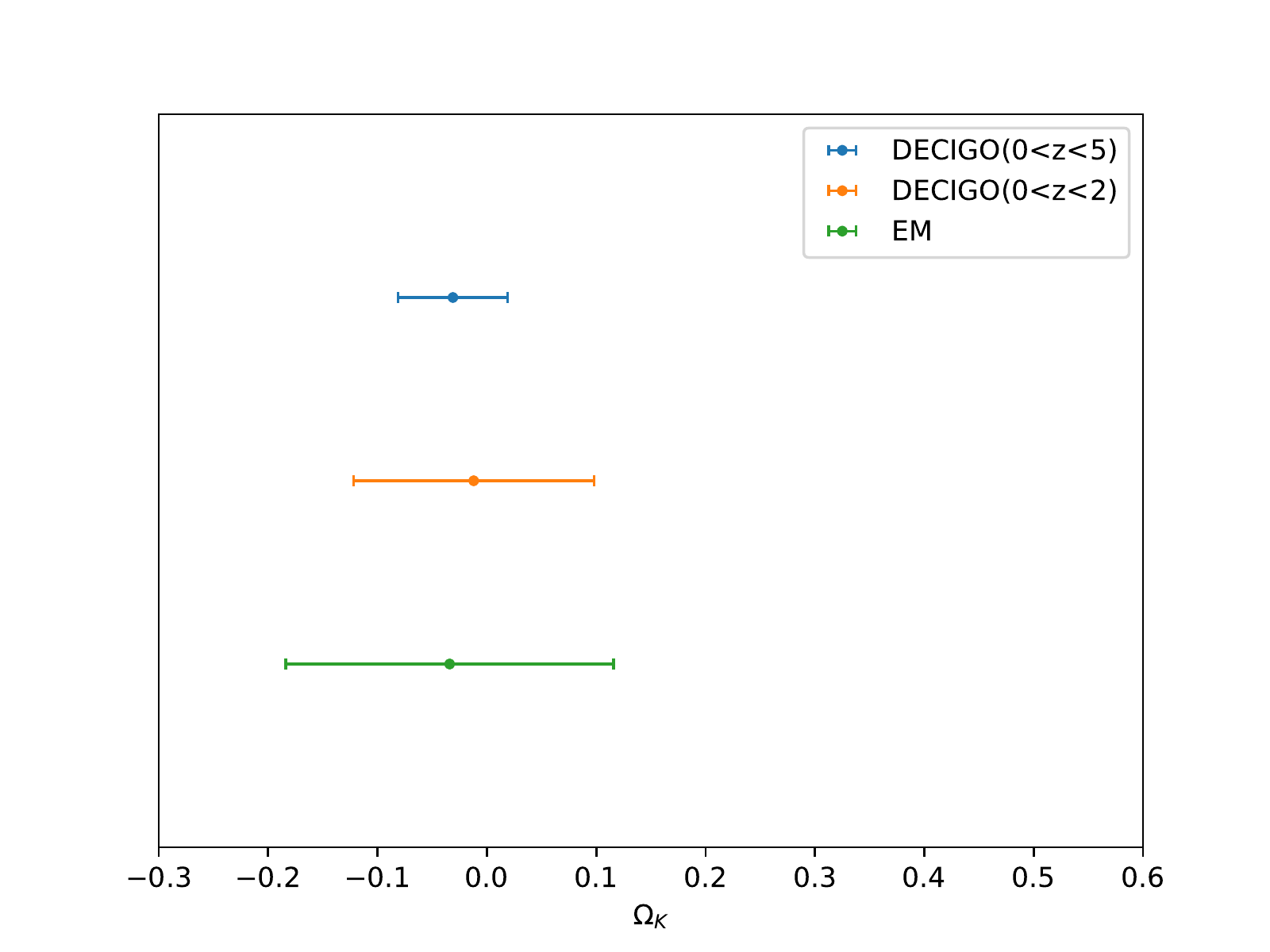}
\caption{\label{fig:threeData} The $\Lambda CDM+\Omega_{K}$ model-dependent curvature density parameter $\Omega_{K}$}
\hfill
\end{figure}

As shown in the Tab. \ref{tab:Omegak2}, The results of constraining $\Lambda CDM+\Omega_{K}$ model in the table show that the curvature best values and their $68\%$ confidence errors are $\Omega_{K}=-0.012\pm0.11$ and $\Omega_{K}=-0.031\pm0.05$ constrained by sub-sample DECIGO data at the redshift $0$ to $2$ and full sample at $z=0$ to $5$ respectively. In addition, when two kinds of EM (supernova + quasar) data are used, the constraint of curvature density parameter is $\Omega_{K}=-0.034\pm0.15$. For intuitive comparison, we also show the model-based constraint results of three kinds of data in Fig. \ref{fig:threeData}.

\begin{table}[tbp]
\centering
\begin{tabular}{c|cccccc}
\hline
Data            & $\Omega_{K}$      &  $\Omega_m$      &$H_0$             \\
\hline
EM              &$-0.034\pm0.15$  &$0.292\pm0.071$   & $71.14\pm0.25$   \\
DECIGO (0<z<2)  &$-0.012\pm0.11$  &$0.284\pm0.046$   & $70.16\pm0.43$   \\
DECIGO (0<z<5)  &$-0.031\pm0.05$  &$0.292\pm0.016$   & $70.24\pm0.40$   \\

\hline
\end{tabular}
\caption{\label{tab:Omegak2} Results of $\Lambda CDM+\Omega_{K}$ models with different data constraints.}
\end{table}

Our results show that in the method of directly assuming cosmological models($\Lambda CDM+\Omega_{K}$), the accuracy of curvature parameter error ($\Delta\Omega_{K}=0.05$) given by the full sample with red shift ($z\in [0,5] $) is higher, which is only $0.5$ times of the curvature error($\Delta\Omega_{K}=0.11$) given by the sub-sample ($z\in[0,2]$), and only $0.3$ times of the error ($\Delta\Omega_{K}=0.15$) given by the EM data sample.
It can be seen that the error given by the full sample ($z\in[0,5] $) is more than twice the result of Planck data constraint $\Delta\Omega_{K}=0.018$ \citep{Aghanim2020}. In addition, it is worth noting that the accuracy of the curvature parameter ($\Delta\Omega_{K}=0.11$) is lower than that of the improved curvature test method ($\Delta\Omega_{K}=0.016$) error given under the same redshift range ($z\in[0,2]$), as shown in Tab. \ref{tab:Omegak1}. However, the curvature constraint results obtained by improved curvature test method are similar to those obtained by Planck data. For EM data, compared with the constraint result of using supernova data alone on the model is $\Omega_{K} = -0.062^{+0.189}_{-0.169}$ \citep{Gao2020}, the accuracy of curvature is improved after adding quasars, but it is still lower than those from GW data. Therefore, for gravitational waves, a improved curvature test method to constrain curvature can be able to obtain higher accuracy results. In addition, for the constraint results of matter density parameters ($\Omega_{m}$) in the $\Lambda CDM+\Omega_{K}$ model, the constraint accuracy of GW data is also higher than that of EM data, and this constraint ability becomes stronger with the increase of the number of events with higher redshift of GW. For the Hubble constant($H_0$), the error results constrained by the two kinds of GW data are slightly larger than those given by EM data, but they are consistent with each other within the $1\sigma$ error range.

\section{Conclusion} \label{Con}

Gravitational waves as a standard siren may open a new window for the study of cosmology, and more interesting results are expected through gravitational wave detection. In this paper, we use a improved curvature test method to study curvature. Firstly, 31 sets of Hubble parameter data from the cosmic chronometer are reconstructed using Gaussian process. After integrating the reconstructed data, a set of luminosity distance $d_L^{Hz}(z)$ given by $H(z)$ is obtained. Then, we use the data from DECIGO to simulate 10000 GW events, and obtain their redshift, luminosity distance and corresponding error. By comparing the luminosity distance given by the two kinds of data, the curvature parameters are constrained. In our work, 1. we simulated 10000 GW events based on the estimation of future detection events by the DECIGO project. 2. The curvature constraint results of the third generation GW detectors (ET) and the space gravitational wave detectors (DECIGO) are compared in our study. When the cosmic curvature is constrained by the improved curvature test method, ET and DECIGO get the constraint results of $\Omega_{K}=0.035\pm0.039$,$\Omega_{K}=-0.007\pm0.016$, respectively. The results show that the space gravitational wave detectors can provide stronger constraint effect. 3. We also compare the results of using improved curvature test method and the method of using $\Lambda CDM+\Omega_{K}$ model to constrain curvature based on DECIGO data, which are $\Omega_{K}=-0.007\pm0.016$, $\Omega_{K}=-0.031\pm0.05$. It shows that for DECIGO, improved curvature test method can get stronger constraint effect, and the curvature constraint accuracy obtained by improved curvature test method can be similar to that of curvature constraint given by Planck2018 microwave background ($\Delta \Omega_{K}=0.018$) \citep{Aghanim2020}.

 At the same time, we also compare the constraint results given by other observation data. Compared with the third-generation gravitational wave detector ET\citep{Wei2018} and supernovae\citep{Gao2020} and quasars\citep{Cao2019b}, the error given by DECIGO is half that of ET, and the accuracy is one order of magnitude higher than that of supernovae and quasars.

In addition, we also use DECIGO to constrain the $\Lambda CDM+\Omega_{K}$ model, and take the EM data as the control group. The results show that in the case of constraint model, the curvature constraint effect of DECIGO is slightly higher than that of EM. Meanwhile, GW have great potential for curvature constraints under improved curvature test method conditions, and by comparing the current constraint results, the gravitational wave detector DECIGO has higher constraint ability of curvature constraints than ET and some other current research results. We expect future gravitational wave detection projects to provide a more accurate constraint on cosmic curvature. At the same time, we note that some scholars have proposed a framework including peculiar velocity correction for GW sources\cite{Mukherjee2021}, which will provide guidance for our next work direction.

\acknowledgments

This work was supported in part by the National Natural Science Foundation of China under Grant Nos. 11873001, 12047564 and by the Fundamental Research Funds for the Central Universities under Grant No. 2020CDJQY-Z003. This project is sponsored by the Scientific Research and Innovation Project of Graduate Students in Chongqing No. CYS20272.


%\bibliographystyle{JHEP}
%\bibliography{DECIGO-Curvature}
% Please avoid comments such as "For a review'', "For some examples",
% "and references therein" or move them in the text. In general,
% please leave only references in the bibliography and move all
% accessory text in footnotes.

% Also, please have only one work for each \bibitem.

\end{document}